\documentclass{revtex4}

\usepackage{graphicx}
\def\vtotbar{\overline V_{\rm tot}}
\def\dbar{\overline D}
\def\del{\delta}

\def\mx{M_X}

\def\mz{m_Z}

\def\h{h}
\def\a{a}
\def\mh{m_{\h}}

\def\lam{\lambda}

\def\hpm{H^{\pm}}

\def\call{{\cal L}}

\def\what{\widehat}
\def\tauptaum{\tau^+\tau^-}

\def\lsim{\mathrel{\raise.3ex\hbox{$<$\kern-.75em\lower1ex\hbox{$\sim$}}}}
\def\gsim{\mathrel{\raise.3ex\hbox{$>$\kern-.75em\lower1ex\hbox{$\sim$}}}}
\def\ifmath#1{\relax\ifmmode #1\else $#1$\fi}

\def\vev#1{\langle #1 \rangle}
\def\lam{\lambda}

\def\mplanck{M_{\rm P}}

\def\etc{{\em etc.}}

\def\eg{{\it e.g.}}

\def\dmm{\Delta^{--}}

\def\mdmm{m_{\dmm}}
\def\hdmm{h^{\dmm}}
\def\dpp{\Delta^{++}}
\def\delp{\Delta^{+}}

\def\hzero{\Delta^0}

\def\gamdmm{\Gamma_{\dmm}^T}

\def\stop{\wt t}

\def\mstop{m_{\stop}}

\def\msusy{m_{\rm SUSY}}

\def\eg{{\it e.g.}}

\def\hsm{h_{\rm SM}}
\def\mhsm{m_{\hsm}}
\def\hl{h^0}
\def\hh{H^0}
\def\ha{A^0}
\def\hp{H^+}
\def\hm{H^-}
\def\hpm{H^{\pm}}
\def\mhl{m_{\hl}}
\def\mhh{m_{\hh}}
\def\mha{m_{\ha}}

\def\mhpm{m_{\hpm}}
\def\tanb{\tan\beta}

\def\mz{m_Z}
\def\mw{m_W}
\def\mgut{M_U}
\def\mx{M_X}

\def\wp{W^+}
\def\wm{W^-}

\def\cnone{\wt\chi^0_1}

\def\snu{\wt\nu}

\def\mcnone{m_{\cnone}}

\def\wt{\widetilde}

\def\cpone{\wt \chi^+_1}
\def\cmone{\wt \chi^-_1}
\def\cpmone{\wt \chi^{\pm}_1}

\def\mcpmone{m_{\cpmone}}

\def\stau{\wt \tau}

\def\emem{e^-e^-}

\def\gamdmm{\Gamma_{\dmm}^T}

\def\MPL #1 #2 #3 {{\sl Mod.~Phys.~Lett.}~{\bf#1} (#3) #2}
\def\NPB #1 #2 #3 {{\sl Nucl.~Phys.}~{\bf #1} (#3) #2}
\def\PLB #1 #2 #3 {{\sl Phys.~Lett.}~{\bf #1} (#3) #2}
\def\PR #1 #2 #3 {{\sl Phys.~Rep.}~{\bf#1} (#3) #2}
\def\PRD #1 #2 #3 {{\sl Phys.~Rev.}~{\bf #1} (#3) #2}
\def\PRL #1 #2 #3 {{\sl Phys.~Rev.~Lett.}~{\bf#1} (#3) #2}
\def\RMP #1 #2 #3 {{\sl Rev.~Mod.~Phys.}~{\bf#1} (#3) #2}
\def\ZPC #1 #2 #3 {{\sl Z.~Phys.}~{\bf #1} (#3) #2}
\def\IJMP #1 #2 #3 {{\sl Int.~J.~Mod.~Phys.}~{\bf#1} (#3) #2}
\def\NIM #1 #2 #3 {{\sl Nucl.~Inst.~and~Meth.}~{\bf#1} {#3} #2}
\def\lam{\lambda}
\def\br{B}
\def\tauptaum{\tau^+\tau^-}

\def\gam{\gamma}

\def\anti{\overline}
\def\epem{e^+e^-}
\def\mupmum{\mu^+\mu^-}
\def\zstar{Z^\star}

\def\mupmum{\mu^+\mu^-}

\def\rts{\sqrt s}
\def\ie{{\it i.e.}}
\def\eg{{\it e.g.}}
\def\eps{\epsilon}
\def\anti{\overline}

\def\fbi{~{\rm fb}^{-1}}

\def\abi{~{\rm ab}^{-1}}

\def\mev{~{\rm MeV}}
\def\gev{~{\rm GeV}}
\def\tev{~{\rm TeV}}

\def\dmm{\Delta^{--}}
\def\mdmm{m_{\dmm}}
\def\hdmm{h^{\dmm}}
\def\dpp{\Delta^{++}}

\def\hzero{\Delta^0}

\def\gamdmm{\Gamma_{\dmm}^T}

\newcommand{\nc}{\newcommand}
\nc{\beq}{\begin{equation}}   \nc{\eeq}{\end{equation}}
\nc{\bea}{\begin{eqnarray}}   \nc{\eea}{\end{eqnarray}}
\nc{\baa}{\begin{array}}      \nc{\eaa}{\end{array}}
\nc{\bit}{\begin{itemize}}    \nc{\eit}{\end{itemize}}
\nc{\ben}{\begin{enumerate}}  \nc{\een}{\end{enumerate}}
\nc{\bce}{\begin{center}}     \nc{\ece}{\end{center}}
\def\beqa{\begin{eqnarray}}
\def\eeqa{\end{eqnarray}}
\def\bed{\begin{description}}
\def\eed{\end{description}}

\def\eg{{\it e.g.}}

\def\dmm{\Delta^{--}}

\def\mdmm{m_{\dmm}}
\def\hdmm{h^{\dmm}}
\def\dpp{\Delta^{++}}
\def\delp{\Delta^{+}}

\def\hzero{\Delta^0}

\def\gamdmm{\Gamma_{\dmm}^T}

\def\emem{e^-e^-}
\def\dmm{\Delta^{--}}
\def\mdmm{m_{\dmm}}

\def\dpp{\Delta^{++}}
\def\delp{\Delta^{+}}

\def\hzero{\Delta^0}

\def\gamdmm{\Gamma_{\dmm}^T}

\def\dmm{\Delta^{--}}
\def\mdmm{m_{\dmm}}
\def\hdmm{h^{\dmm}}
\def\dpp{\Delta^{++}}

\def\hzero{\Delta^0}

\def\gamdmm{\Gamma_{\dmm}^T}

\def\rta{\rightarrow}
\def\tanb{\tan\beta}

\begin{document}
% You should use BibTeX and revtex.bst for references
\bibliographystyle{revtex}

% number  and your conference paper identification number on the
% title page in preprint mode. Multiple \preprint commands are allowed.
%\preprint{}

\title{Phenomenology of Generalized Higgs Boson 
Scenarios~\footnote{Submitted to the Proceedings
of ``The Future of Particle Physics'', Snowmass 2001, P1 group.}
\footnote{This work was supported in part by the U.S. Department of Energy.}}
% Optional argument for running titles on pages
%\title[]{}



\author{John F. Gunion}
%\email[]{jfgucd@higgs.ucdavis.edu}
%\homepage[]{higgs.ucdavis.edu/gunion/home.html}
%\thanks{This work was supported in part by the U.S. Department of Energy.}
\affiliation{Department of Physics, University of California, Davis, CA 95616}


%\date{\today}

\begin{abstract}
I outline some of the challenging issues that could arise in attempting
to fully delineate various possible Higgs sectors, with focus on
the minimal supersymmetric model,  
a general two-Higgs-doublet model, and extensions thereof.
\end{abstract}
% \pacs{}

\hbox to \hsize{
%
%$\vcenter{
%\hbox{\fortssbx University of California - Davis}
%}$
%
\hfill
$\vcenter{\normalsize
\hbox{\bf UCD-2001-12} 
\hbox{\bf hep-ph/0110362}
\hbox{October, 2001}
%\hbox{Preliminary Version: \today}
}$
}

%\maketitle must follow title, authors, abstract and \pacs
\maketitle

\noindent In a broad sense the three basic topics of this review will be: 
(i) extended Standard Model Higgs sectors; 
(ii) perturbations of `Standard' minimal supersymmetric model (MSSM) 
Higgs sector phenomenology; 
and (iii) Higgs phenomenology for SUSY models beyond the MSSM.
Also interesting are Higgs-like particles and how their phenomenology
would differ from or affect that for Higgs bosons.  Such particles
include: radions; top-condensates; and the pseudo-Nambu-Goldstone bosons of
technicolor.  However, I will not have space to discuss these latter objects.

\section{Extended SM Higgs Sectors}
\label{esm}
Even within the SM context, one should consider extended Higgs sector
possibilities. 
\bit 
\item
One can add one or more singlet Higgs fields.
This leads to no particular theoretical problems (or benefits) 
but Higgs discovery can be much more challenging.
\item One can consider more than one Higgs doublet field, the simplest
case being the general two-Higgs-doublet model (2HDM).
A negative point is that, in the general case, 
the charged Higgs mass-squared is
not automatically positive (as required to avoid breaking
electromagnetism). A more complete model context might, however,
lead to a 2HDM Higgs sector having $\mhpm^2>0$ as part of an effective
low-energy theory.
A positive point is that CP violation can arise
in a Higgs sector with more than one doublet
and possibly be responsible for all CP-violating phenomena.
\item One can include Higgs triplet fields.
If there is a neutral member of the triplet representation
and it has a non-zero vev, then $\rho=\mw/(\mz\cos\theta_W)$ is no 
longer computable \cite{Gunion:1991dt} 
(even if representations and vevs are chosen so that $\rho=1$
at tree level); $\rho$ becomes
another input parameter to the theory. It is not clear how negatively
one should regard this loss of predictability.
Of course, if the neutral vev $=0$, then 
there is no impact on EWSB and $\rho=1$ remains natural.
\item As regards higher representations, we should not forget
that there are special choices of $T$ and $Y$
for the Higgs multiplet, the next simplest after $T=1/2,|Y|=1$
being $T=3,|Y|=4$, that yield $\rho=1$
at tree level and finite loop corrections to $\rho$ even if
the neutral field has non-zero vev \cite{hhg}.
\eit
Coupling constant unification is also an important ingredient
in evaluating the attractiveness of an extended SM Higgs sector.
Let us denote by $N_{T,Y}$ the number of Higgs representations
of weak-isospin $T$ and hypercharge $Y$. 
It is easy to show \cite{Gunion:1998ii} that certain choices
of the $N_{T,Y}$'s can yield coupling constant unification
for SM matter content (i.e. no SUSY), although
not at as high a scale as the standard $\mgut\sim 10^{16}\gev$.
For example,  $N_{{1\over 2},1}=2,N_{1,0}=1$ yields
$\alpha_s(\mz)=0.115$ and $\mgut=1.6\times 10^{14}\gev$.
In the context of extra dimensions, it may be that unification
should occur at scales not so far above a $\tev$. An example
in the SM extension context with low unification scale is
$N_{{1\over 2},1}=N_{{1\over 2},3}=N_{1,2}=N_{1,0}=4,N_{3,4}=3$,
which leads to
$\alpha_s(\mz)=0.112$, $\mgut=1000\tev$ and $\alpha_U=0.04$.
Even lower values of $\mgut$ are possible in the case of 
MSSM matter content (\ie\ including superpartners of the SM particles). 
For example, if $N_{{1\over 2},1}=N_{1,2}=N_{1,0}=4,N_{3,4}=4$ one obtains
$\alpha_s(\mz)=0.114$, $\mgut=4\tev$, and $\alpha_U=0.07$.
So, perhaps one should not discard complicated Higgs sectors out of hand.

Even the simplest extensions of the Higgs sector
can lead to dramatic changes in our ability to detect Higgs bosons.
Discovery prospects will often depend strongly on the type
of collider, although given sufficient $\rts$ it is usually the case that an
$\epem$ collider is the best option.

Current data provide some important hints and constraints regarding
the Higgs sector \cite{lepewwg}.  As is well known, the simplest interpretation
of the precision electroweak data is the existence of a rather
light SM-like Higgs boson (the preferred mass being below the LEP
experimental lower limit of 114 GeV). The $1-CL$
plots as a function of SM-like Higgs boson mass show
that it is also possible to interpret the LEP2 data as being due
to a spread-out Higgs signal, \eg\ several Higgs bosons in the $<114\gev$
region, each with an appropriate fraction of the SM $ZZ$ coupling.
Such a situation was considered in \cite{Espinosa:1999xj} (see also \cite{Akhoury:2001hw} and references therein).
The simplest Higgs sector for which this could occur is one
obtained by adding a modest number of singlet Higgs fields
to the minimal one-doublet SM Higgs sector. For an appropriate
Higgs potential that mixes the many neutral fields, the
physical Higgs bosons would be mixed states sharing the $WW/ZZ$
coupling strength squared and having decays to a rather
confused set of final states. If these Higgs bosons had masses
spread out every $10-20\gev$ (\ie\ smaller than the 
detector resolution in a typical decay channel),
a broad/diffuse `continuum' Higgs signal would be the result.
Fortunately, the constraints outlined
below imply that a future $\epem$ linear collider with $\rts\sim 500\gev$
would be guaranteed to detect even such a signal for currently
anticipated integrated luminosities. In particular, if we define
$
\vev{M^2}\equiv \sum_i C_i^2m_{h_i}^2\,,
$
where $C_igm_W$ is the strength of the $h_iWW$ coupling ($\sum_i C_i^2=1$
being required by unitarity),
precision electroweak data requires that $\vev{M^2}\lsim (200-250\gev)^2$.
(In the SUSY context, even allowing for the most general Higgs representations,
RGE evolution starting with perturbative couplings at $\mgut$ 
implies this same result.)
This sum rule implies \cite{Espinosa:1999xj}
that enough of the resulting diffuse excess
in the recoil $\mx$ mass distribution for $\epem\to ZX$ ($Z\to\epem,\mupmum$)
would be confined to the $\mx\in[100,200\gev]$ mass region
that it could be detected with $L\gsim 200\fbi$ accumulated luminosity
at $\rts=500\gev$, despite the significant background~\cite{Gunion:1998jc}.
It is far from clear that such a no-lose theorem can be established
for this scenario at the LHC.
In particular, the
$\gam\gam$ decay width is reduced (due to less $W$ loop
contribution) for each of the overlapping Higgs states. Meanwhile,
the $W\h$ and $Z\h$ production processes would be weak 
for each of the individual Higgs bosons, $\h_i$, and these $\h_i$ 
signals would be spread out and overlapping in mass. The
$t\anti t\h_i$ signals might retain a roughly SM-like rate
for each $\h_i$, but again the signals would be
spread out by experimental effects and overlapping, so that search
techniques using bump hunting could not be employed.

A more popular SM Higgs sector extension is the general two-Higgs-doublet
model (2HDM) with its five physical Higgs bosons
($h_{1,2,3}^0$, $\hpm$, or in the CP-conserving (CPC) case $\hl,\hh,\ha,\hpm$).
An interesting question is whether or not we are guaranteed to
find at least one of the Higgs bosons of a general 2HDM given
current precision electroweak constraints.  The answer is
yes, but direct Higgs detection might only be possible at the LHC
if the LC has $\rts\lsim 1\tev$. One case in which this 
situation arises is \cite{Chankowski:2000an} if the
only light Higgs boson has no $WW/ZZ$ couplings and all the
other Higgs bosons have mass $\gsim 1\tev$. As we describe
shortly, these heavy Higgs bosons can be chosen to have mass splittings
such that the $S,T$ parameters
fall within the current 90\% precision electroweak ellipse.

At the LC, the relevant discovery processes for a $\h$ with no $WW,ZZ$
couplings are: $\epem\to t\anti t \h$ and 
$\epem\to b\anti b \h$~\cite{Grzadkowski:2000wj}; 
$\epem\to Z^*\to  Z \h\h$ \cite{Haber:1993jr,gunfar}; 
$\epem\to\epem W^*W^*\to\epem  \h\h$~\cite{Djouadi:1996jf,gunfar};
$\gam\gam\to \h$~\cite{gunasner}. 
That these processes might have reasonable rates follows from
the couplings involved.
It can be shown \cite{Grzadkowski:1999ye,Grzadkowski:2000wj} 
that for any $\h$ with no (or very small)
$WW/ZZ$ coupling  (both $\h=\ha$ and $\h=\hl$ are possibilities in
the CPC case) at least one of the couplings $t\anti t\h$ or $b\anti b\h$
must be substantial:
$(\hat{S}^t_{\h})^2 + (\hat{P}^t_{\h})^2=\cot^2\beta 
\,,\quad
(\hat{S}^b_{\h})^2 + (\hat{P}^b_{\h})^2 
=\tan^2\beta$,
where $\hat S$ and $\hat P$ are the $1$ (\eg\ $\h=\hl$) and $\gamma_5$ 
(\eg\ $\h=\ha$) 
couplings defined relative to usual SM-type weight. 
The quartic couplings,
$ZZ\h\h$ and $\wp\wm\h\h$, arise from the gauge covariant
structure  $(D_\mu \Phi)^\dagger (D^\mu \Phi)$ and are 
of guaranteed magnitude. The
$\gam\gam\to\h$ coupling derives from fermion loops, and above
we saw that not both the $b\anti b\h$ and $ t\anti t \h$ coupling
can be suppressed. Unfortunately, despite these guarantees
for non-suppressed couplings, there
is a wedge-like region of parameter space where discovery of
a light decoupled $\h$ will be quite difficult.

Turning first to $t\anti t\h$ and $b\anti b\h$ production,
the former (latter) always yields significant rates if $\tanb$ is small (large)
enough (and the process is kinematically allowed),
while $\epem\to b\anti b\h$ always works if $\tanb$ is large enough. 
But, even for high $\rts$ and $L=1000\fbi$,
there remains a wedge of moderate $\tanb$ for which 
neither process provides adequate event rate.
The wedge expands to lower and higher $\tanb$ values as $\mh$ increases.
The corresponding wedge at the LHC is even larger.
For the lower values of $\mh$, double Higgs production via the quartic
couplings will allow discovery at the LC even in the wedge region.
For instance, the process $\epem\to Z^*\to Z\h\h$ gives
a decent event rate for 
$\mh\lsim 150\gev$ ($\mh\lsim 250\gev$) for $\rts=500\gev$ ($\rts=800\gev$),
while $WW\to \h\h$ fusion production probes $\mh\lsim 200$ ($\mh\lsim 300$).
A careful assessment of backgrounds is required to ascertain just
what the mass reach of these processes actually is.

\begin{figure}[t!]
\leavevmode
\begin{center}
\includegraphics[width=6in]{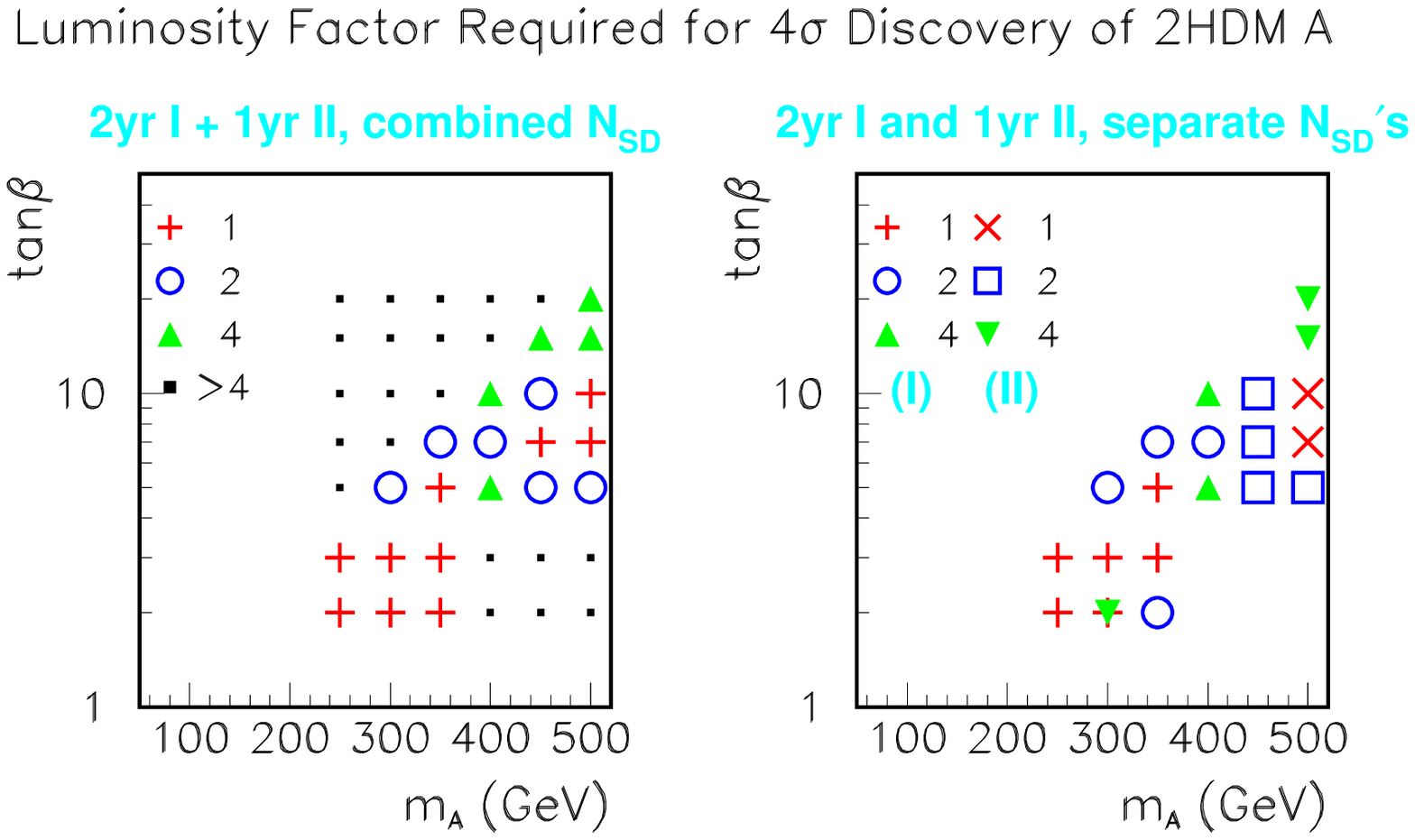}
\end{center}
\vspace*{-3in}
\caption[0]{\label{wedgeplota2hdm}Assuming a machine energy of $\rts=630\gev$,
we show the $[\mha,\tanb]$ points for which two $10^7$ sec years
of $\gam\gam$ collision operation 
using the type-I (broad) $E_{\gam\gam}$ spectrum
and one $10^7$ sec year using the type-II peaked spectrum
configuration will yield $S/\sqrt{B}\geq4$ for
the $\ha$ of a general 2HDM, assuming all other 2HDM
Higgs bosons have mass of $1\tev$.  In the left-hand window we
have combined results from the type-I and type-II running using
$S/\sqrt B=\sqrt{ S_I^2/B_I+S_{II}^2/B_{II}}$. In the right-hand
window we show the separate results for $S_I/\sqrt{B_I}$ and 
$S_{II}/\sqrt{B_{II}}$. Also shown are the additional points for
which a $4\sigma$ signal level is achieved if the total
luminosity is doubled or quadrupled (the `2' and `4' symbol cases)
relative to the 2+1-year luminosities we are employing.
(In the LH window, the small black squares indicate the additional
points sampled for which even a luminosity increase of a factor
of 4 does not yield a $4\sigma$ signal.) 
Such luminosity
increases could be achieved for some combination of longer running time and/or
improved technical designs. For example, the factor of `2' 
results probably roughly apply to TESLA.   
Cuts and procedures are as described in \cite{gunasner}.
}
\end{figure}

If the $\gam\gam$ collider option is implemented at the LC,
$\gam\gam\to \h$ will provide a signal for a decoupled $\h$
over a significant portion of the wedge region. The
results from the quite realistic study of \cite{gunasner} are
illustrated in Fig.~\ref{wedgeplota2hdm}, which focuses on
the case of $\h=\ha$ and $\mha\geq 250\gev$. The crosses and
pluses indicate $4\sigma$ discovery points after 3 years of appropriate
running at the NLC. The higher TESLA luminosity for $\gam\gam$ collisions
would allow $4\sigma$ discovery for the additional points indicated
by the circles and squares.

Finally, although we don't present details here, a muon collider would
probably be able to provide $4\sigma$ signals for any $\h$ in the $\mh<500\gev$
wedge region after about 3 years of appropriately configured operation,
assuming the nominal Higgs factory luminosities discussed during this
workshop. For more details, see \cite{jfgucla}.

\begin{figure}
\begin{center}
\includegraphics[width=12cm]{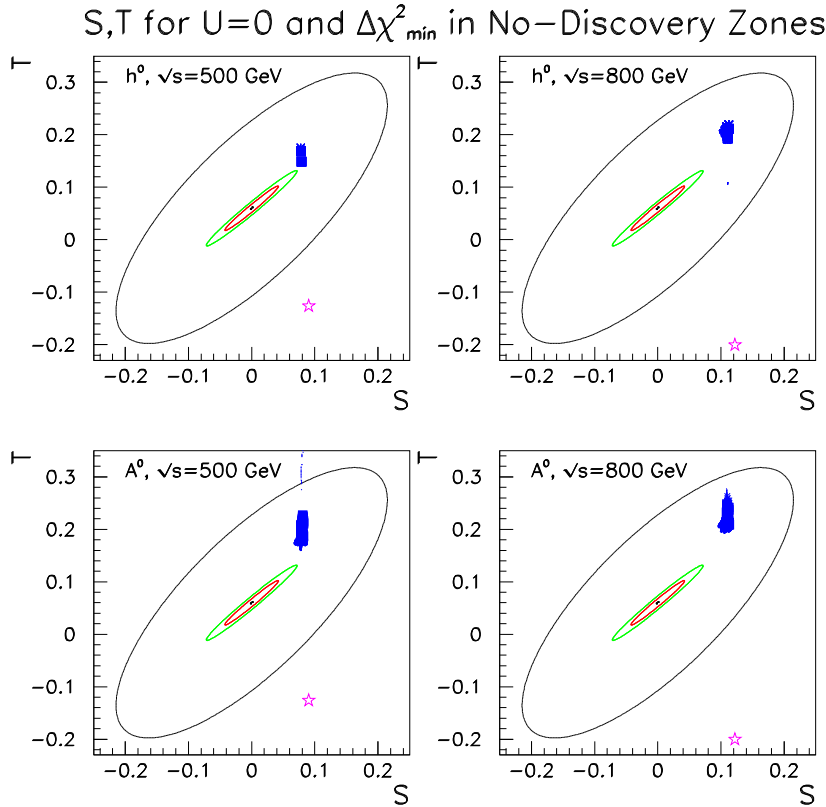}
\caption{\label{ellipse}
The outer ellipses show the 90\% CL region from current
precision electroweak data in the $S,T$ plane
for $U=0$ relative to a central point defined
by the SM prediction with $\mhsm=115$ GeV. 
The blobs of points show the $S,T$ predictions for 2HDM  models 
with a light $\hl$ or $\ha$ with no $WW/ZZ$ coupling and with
$\tanb$ such that the light $\hl$ or $\ha$ cannot be detected
in $b\anti b$+Higgs or $t\anti t$+Higgs production
at either the LC or the LHC; the mass of
the SM-like  Higgs boson of the model is set equal to $\rts=500\gev$ (left)
or $800\gev$ (right) and the heavier Higgs masses have been
chosen to minimize the $\chi^2$ of the full precision electroweak fit.
The innermost (middle) ellipse shows the  90\% (99.9\%) CL
region for $\mhsm=115$ GeV after
Giga-$Z$  LC operation {\it and} a $\Delta m_W\protect\lsim 6$ MeV threshold
scan measurement. The stars to the bottom right show the $S,T$ predictions
in the case of the SM with $\mhsm=500\gev$ (left) or $800\gev$ (right).} 
\end{center}
\end{figure}

Is the type of scenario being considered (a light decoupled $\h$
and all other Higgs bosons heavy) consistent with precision electroweak
constraints? In fact, it can be arranged \cite{Chankowski:2000an}.
For example, consider the case of
$\h=\ha$ and  a SM-like $\hl$ with mass $\sim 1\tev$. The heavy 
$\hl$ leads to large 
$\Delta S>0$ and large $\Delta T<0$ contributions, which 
on its own would place the $S,T$ prediction of the 2HDM model 
well outside the current 90\% CL ellipse --- see the stars in 
Fig.~\ref{ellipse}.
However, large $\Delta T<0$ contribution from the SM-like $\hl$
can be compensated by a large $\Delta T>0$ from a
small mass non-degeneracy (weak isospin breaking) of the
still heavier $\hh$ and $\hpm$ Higgs bosons. In detail, for a light $\ha$
one finds
\beq
   \Delta \rho=\frac{\alpha}{16 \pi m_W^2 c_W^2}\left\{\frac{c_W^2}{s_W^2}
   \frac{m_{H^\pm}^2-m_{H^0}^2}{2}-3m_W^2\left[\log\frac{m_{h^0}^2}{m_W^2}
   +\frac{1}{6}+\frac{1}{s_W^2}\log\frac{m_W^2}{m_Z^2}\right]\right\}\nonumber
\label{drhonew}
\eeq
from which we see that the first term can easily compensate
the large negative contribution to $\Delta\rho$ from the $\log (\mhl^2/\mw^2)$
term. The overall arrangement 
of this light $\ha$ case is further illustrated in 
the lower windows of Fig.~\ref{ellipse}.
The blobs correspond to 2HDM parameter choices for which
$\mhl=\rts$ (so that it cannot be observed at the LC of this $\rts$)
and $\mhpm-\mhh\sim {\rm few}\gev$ has been chosen
(with both $\mhpm,\mhh\sim 1\tev$)
so that the $S,T$ prediction is well within the 90\% CL ellipse,
while at the same time $\mha$ and $\tanb$ are precisely 
in the wedge of parameter space
for which the LHC and $\epem$ LC operation
would not allow discovery of the $\ha$. However, this scenario
can only be pushed so far. In order to maintain perturbativity
for all the Higgs self couplings, it is necessary that $\mhl\lsim 1\tev$,
implying that it would be detected at the LHC. 
Giga-$Z$ operation and a $\Delta\mw=6\mev$ $WW$ threshold scan at the LC
(with the resulting ellipse sizes illustrated in Fig.~\ref{ellipse}) 
would be very important to confirm that the $S,T$ values were 
indeed those corresponding
to a $S,T$ location like that of the blobs of Fig.~\ref{ellipse}. If no
other new physics was detected at the LC or LHC that could cause
the extra $\Delta T>0$, searching for a possibly light decoupled $\ha$ 
would become a high priority. 

\begin{figure}[h]
\begin{center}
\includegraphics[width=10cm]{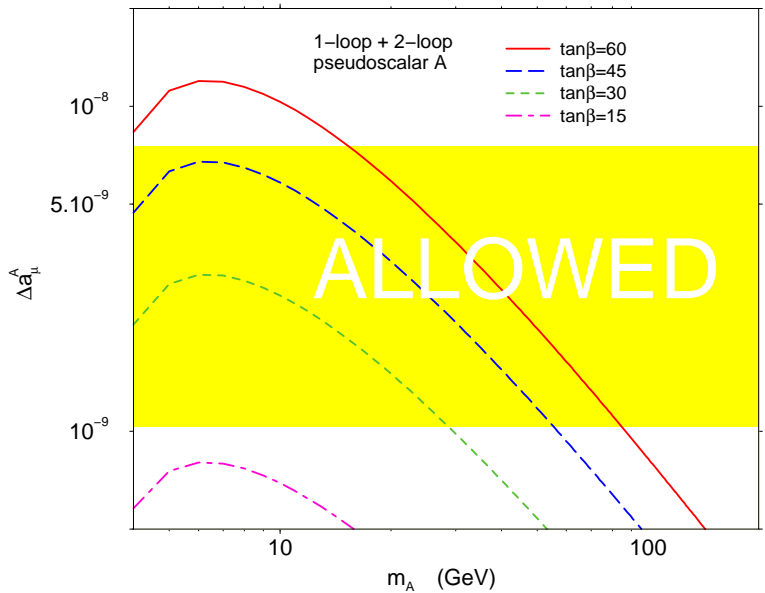}
\caption{\label{f:amu} Explanation of new BNL $a_\mu$ value 
via a light 2HDM $\ha$, from~\cite{Cheung:2001hz}.}
\end{center}
\end{figure}

Interestingly, the recent observation \cite{Brown:2001mg} of
a discrepancy ($10.3\times 10^{-10} < \Delta a_\mu < 74.9 \times 10^{-10}$ 
at 95\% C.L. ($\pm 1.96\sigma$) for 
`standard' $\sigma(\epem\to \mbox hadrons)$ at low $\rts$)
with SM predictions for  $a_\mu$ can be explained by 
the existence of a light $\ha$ \cite{Cheung:2001hz}.
A light $\ha$ ($\hl$) gives a 
positive (negative) contribution to $a_\mu$, dominated
by the two-loop Bar-Zee graph. 
As shown in Fig.~\ref{f:amu}, rather small values of $\mha$ and large values of
$\tanb$ are needed to explain the entire $\Delta a_\mu$.
In the indicated range of $\tanb>17$, the $\ha$ 
will be found at the LC for sure
and possibly  also at the LHC. However, it seems possible
that the $\Delta a_\mu$ discrepancy will turn out
to be not quite so large as currently stated, 
either as statistics improve or because the forthcoming
low-$E$ $\sigma(\epem\to {\rm hadrons})$ data alters the SM prediction. 
Smaller values for $\Delta a_\mu$ would be best explained by
smaller $\tanb$ and  higher 
$\mha$ values that could lie inside the LC/LHC no-discovery wedge region.

Extra dimensions and related ideas can have a tremendous impact
on Higgs phenomenology.  There is only space for the most cursory of
reviews. In the simplest model, 
SM particles live on a `brane' (3+1 dimensions), and gravity
resides in the bulk \cite{Arkani-Hamed:1998rs,Antoniadis:1998ig}. 
The new physics scale, $\Lambda$, typically identified with
the string scale, $M_S$, is possibly as small as $1\tev$.
Since the quadratic divergence at 1-loop for $\mhsm^2$ is cutoff by 
the string physics at $M_S$, a light Higgs boson would be
natural in the SM.
Small fermionic couplings could arise if the brane is `fat'
and the fermion fields (other than the top) 
are localized within the brane so as to have little
overlap with the Higgs field(s) \cite{Mirabelli:2000ks}.
Some important results of these ideas are the following.
\bit
\item
Extra contributions to precision electroweak parameters from
effective operators proportional to $1/M_S$ can be substantial,
and in a fashion somewhat analogous to the 2HDM discussion
yield an extra positive $\Delta T$ contribution that would
allow for the SM Higgs boson to be heavy~\cite{Hall:1999fe,Rizzo:2000br}.
As in the 2HDM case, $\mhsm\lsim 1\tev$ is required and 
other signals of the extra dimensional physics would emerge
at energy scales near a TeV.
\item
The KK graviscalar excitations could provide the mechanism for
electroweak symmetry breaking~\cite{Grzadkowski:2000xh}.
In the simple case studied, all SM particles live on the brane.
One must minimize an effective potential consisting of $V(\phi)- 
\call_{\rm mass}(\phi_{KK}^{\vec n})-\call_{\rm mix}(\phi_{KK}^{\vec n},\phi)$,
where $\phi$ is the usual Higgs field,
 $\call_{\rm mass}$ contains the quadratic mass terms
for the KK graviscalar fields $\phi_{KK}^{\vec n}$, and 
$\call_{\rm mix}\propto \kappa\sum_{\vec n}\phi_{KK}^{\vec n} T_\mu^{\mu,~\rm Higgs}
\propto\kappa \sum_{\vec n}\phi_{KK}^{\vec n}V(\phi)$
arises because gravity sees the energy-momentum tensor. Although
$\kappa\propto 1/\mplanck$ is small, there are many $KK$ modes.
After integrating out $KK$ modes, one finds $\vtotbar=V(\phi)-\dbar V^2(\phi)$,
where 
$\dbar\equiv \kappa^2{\del-2\over \del+2}
\sum_{{\rm all}\,\vec n}{1\over m_n^2}$ ($\delta=$number of extra dimensions).
For $\delta>2$,
the sum is divergent -- after regulation by the string,
$\dbar\sim M_S^{-4}$ with a coefficient whose sign depends upon the string 
regulation.  It is possible that $\dbar<0$.
Note that even if $V(\phi)={1\over 2}m^2\phi^2+\Xi$ 
(\ie\ no quartic self interactions), 
$\call_{\rm mix}$ generates $\phi^4$ interactions (of correct sign if
$\dbar<0$).
If $\dbar<0$, then $\vtotbar$ has a minimum at $V(\phi)={1\over 2\dbar}$,
which determines values for $\phi$ and the $\phi_{KK}$ fields at the minimum.
Expanding about the vev's, rescaling $\phi\to\what\phi$
for canonical normalization, and diagonalizing the mass
matrix, one finds: a Higgs boson $s_{\rm phys}$ with 
$m_{s_{\rm phys}}^2>0$;
standard $WW/ZZ$ couplings for $s_{\rm phys}$ (with tiny corrections);
no fermionic couplings of $s_{\rm phys}$ at tree level;
and large decays of $s_{\rm phys}$ to states containing two 
graviscalar $KK$ excited states (which are invisible decays).
\item There is a possibility (for a normal EWSB minimum) of large mixing
between graviscalar-KK excitations
and the SM Higgs that could lead to an
effectively invisible Higgs boson~\cite{Giudice:2001av}. 
For this, one must
introduce a $-{\zeta\over 2}R(g)\phi\phi^\dagger$ interaction, 
where $R$ is the usual Ricci scalar.
This interaction leads to an addition to $T_\mu^\mu$ for the $\phi$:
in unitary gauge
$\Delta T_\mu^\mu=-6\zeta vm_h^2h$ (where $h$ is the usual physical
Higgs boson eigenstate in the absence of mixing) and the graviscalar
KK modes $\phi_{KK}^{\vec n}$ couple to this:
$\call\ni {f(\del)\over \mplanck}\sum_{\vec n}\phi_{KK}^{\vec n}T_\mu^\mu$.
The resulting $h$--$\phi_{KK}^{\vec n}$ mixing must be removed
by rediagonalization, and the physical Higgs ends up having
some (invisible) KK-graviscalar excitation components and KK pair decay modes.

\eit

In fact, there are many models 
in which the SM Higgs decays invisibly. 
(Aside from extra dimension models
discussed above, there are models with invisible Majoron decays
and the like \cite{Diaz:1998zg}.)
Thus, it is important to assess discovery prospects for
an invisibly decaying Higgs boson.
This has been studied for various colliders by many people.
I give a very brief summary.
At LEP2 or the LC, one simply looks for $\epem\to ZX$.
For any Higgs with $ZZ$ coupling, the 
recoil $\mx$ distribution will show a peak. 
The LEP2 limit on a single Higgs with SM-like coupling to $ZZ$ derived by
looking for excess $\epem\to ZX$ events is $\mh\geq 114\gev$ \cite{lephiggs},
\ie\ essentially at the kinematic 
limit, even after allowing for the most general mixture
between normal and invisible decay modes.
The LC discovery potential for an invisibly decaying $\h$
with SM-like $ZZ$ coupling would presumably also approach 
the $ZX$ kinematic threshold.
What is possible for a 
Higgs with only fermionic couplings that decays invisibly has not 
been studied in the LC specific context.  Presumably, $\epem\to Z \h\h$.
$t\anti t\h$ and $b\anti b\h$ (all of which provide an event trigger
of visible plus missing energy) would all be useful.
Discovery of a $\h$ with SM-like $WW/ZZ$ couplings that decays invisibly
is more difficult at hadron colliders than at an LC.
One  would employ $W \h$, $Z \h$ 
production \cite{Frederiksen:1994me,Choudhury:1994hv}
or $WW\to\h$ fusion (with jet tagging) \cite{Eboli:2000ze}.
At the Tevatron ~\cite{Martin:1999qf},
it will take $L>5\fbi$ of integrated luminosity just to surpass 
the LEP2 limit. At the LHC with $L=100\fbi$, $W\h,Z\h$ production
 will probe up to $\mh\sim 200\gev$;
in $WW$ fusion, the estimated reach is 300 -- 500 GeV.
For any $\h$ with SM-like $t\anti t$ coupling, 
$t\anti t \h$ production will provide a good signal
at the LHC for $\mh\lsim 250-300\gev$ assuming 
$L=100\fbi$~\cite{Gunion:1994jf}.
Of course, this latter mode, which relies on
the $t\anti t\h$ coupling, is complementary to
the $W\h$ and $Z\h$ modes that rely on the
$VV\h$ coupling.  Further work on both is desirable.

There is no space to more than briefly mention
Higgs triplet models.  Higgs triplet representations with $|Y|=2$ are an
integral part of any left-right symmetric model (LRM) in which
neutrino masses arise via the see-saw mechanism.
Basic collider phenomenology for such models is studied in
\cite{Deshpande:1991ip,Gunion:1989in,Huitu:1997su,Huitu:1999qx}.
The $2\times 2$ notation for the $|Y|=2$ Higgs triplet fields is
$\Delta=\pmatrix{\delp/\sqrt{2} & \dpp \cr \hzero & -\delp/\sqrt{2} \cr}\,.$
The most important new aspect of a Higgs triplet
model (HTM) is the lepton-number-violating coupling:
$
{\cal L}_Y=ih_{ij}\psi^T_{i} C\tau_2\Delta\psi_{j}+{\rm h.c.}
\,,\quad i,j=e,\mu,\tau\,,
$
which, among other things,
 leads to $\emem\rta\dmm$ and $\mu^-\mu^-\to \dmm$ couplings.
Limits on the $h_{ij}$ by virtue of the $\dmm\rta
\ell^-\ell^-$ couplings are best expressed by writing
$|\hdmm_{\ell\ell}|^2\equiv c_{\ell\ell} \mdmm^2(\gev)\,.$ 
A pre-1999 summary of these limits can be found in
 \cite{Gunion:1996mq,Huitu:1999qx}. The strongest
of these limits are (there are no limits on $c_{\tau\tau}$): 
$c_{ee}< 10^{-5}$ (Bhabbha); 
$c_{\mu\mu}<5\times10^{-7}$ ($(g-2)_\mu$ -- I have updated
this limit to reflect the BNL $a_\mu$ data ---
the predicted contribution has the wrong sign); and 
$\sqrt{c_{ee}c_{\mu\mu}}<10^{-7}$ (muonium-antimuonium).
The most likely case (advocated in \cite{Gunion:1996mq}) is that 
$\vev{\Delta^0}=0$, in which case $\rho=1$ remains
natural \cite{Gunion:1991dt}. (In the LRM, the $\Delta^0$
with zero vev would be the neutral member of the
`left' triplet. It would be the members of this `left' triplet
to which the ensuing discussion applies. The `right' triplet
has very different phenomenology.) 
For $\vev{\Delta^0}=0$, the total width $\gamdmm$  would be small and 
large $s$-channel $e^-e^-$ and $\mu^-\mu^-$ production rates are predicted.
The strategy would be as follows. One would first
discover the $\dmm$ in
$p\anti p\to \dmm\dpp$ with $\dmm \to
\ell^-\ell^-,\dpp\to\ell^+\ell^+$ ($\ell=e,\mu,\tau$) at the upgraded
Tevatron or at the LHC \cite{Gunion:1996pq} if $\mdmm\lsim 1\tev$.
According to how it decays -- 
$\dmm\to e^-e^-$, $\mu^-\mu^-$, or $\tau^-\tau^-$ --
we would know which lepton violating couplings are most substantial.
Then, if $\mdmm$ is in a mass range accessible
to the LC or a muon collider,
it would be of paramount importance to build the relevant
$e^-e^-$ and/or $\mu^-\mu^-$ colliders and study $s$-channel production
of the $\dmm$ in order to determine
the actual size of these couplings. For $c_{\ell\ell}$ near current
upper limits,
event rates would be enormous \cite{Gunion:1996mq,Gunion:1998ii};
equivalently one can probe very small $c_{\ell\ell}$ --- at least
a factor of $10^8-10^9$ improvement over current limits would be possible.
Most importantly, if the magnitudes of $c_{ee}$
and $c_{\mu\mu}$ are
such as to be relevant to neutrino mass generation,
observation of $\dmm$ in $s$-channel $e^-e^-$
and $\mu^-\mu^-$ production
would be possible and would allow an actual measurement of these
very fundamental couplings. 

\section{Higgs Bosons in Supersymmetry}

Supersymmetry remains the most attractive solution to
the naturalness and hierarchy problems. Further, the MSSM implies
coupling constant unification at $\mgut\sim {\rm few}\times 10^{16}\gev$
and generates EWSB automatically
via RGE evolution from $\mgut$ beginning with universal 
soft-supersymmetry-breaking masses. These very attractive
features argue strongly for the MSSM model or 
the simplest generalizations thereof that maintain
its attractive features. Overall, it is clearly important to consider
the discovery and study of Higgs bosons in the SUSY context \cite{hhg}.

The MSSM contains exactly two doublets ($Y=+1$ and $Y=-1$),
as required to give masses to both up and down quarks.
Two doublets are also required in order that the 
anomalies generated by the higgsino partners
of the Higgs bosons cancel. Two doublets (and any number of 
singlets) yield perfect coupling constant unification if
the SUSY scale is $\msusy\sim 1\tev$. (Actually, significant
MSSM matter superpartner content at $10\tev$ is advantageous
for obtaining  $\alpha_s(\mz)<0.12$.)
More doublets, triplets, \etc\ would imply a need for intermediate-scale
matter between the TeV and $\mgut$ scales in order to
achieve coupling constant unification.
But, if there are extra dimensions, unification at $\mgut$ may be irrelevant!
As is well known, there are strong 
theoretical bounds on $\mhl$ deriving from the structure of the MSSM.  
(In discussing these bounds, we will take $\mstop\leq 1\tev$, but one should 
keep in mind
the earlier remark regarding some motivation for sparticle masses that
are much higher.)

In the two-doublet MSSM, $\mhl\lsim 130-135\gev$ is predicted,
 although extra dimension effects might allow additional flexibility.
Adding singlets, as in the NMSSM \cite{Ellis:1989er}
(where one complex Higgs singlet 
field is added), relaxes this upper bound on $\mhl$ to
roughly $140\gev$ \cite{Ellwanger:1999ji}, {\it assuming perturbativity
for the new coupling(s) up to $\mgut$}.
Adding more doublets lowers the upper bound.
Adding the most general structure ($Y=2$ triplets being the `worst'
for moving up the mass bound), and allowing the most general mixings \etc,
one finds (assuming perturbativity up to $\mgut$) an upper
bound of $\sim 200\gev$~\cite{Espinosa:1998re}.

Experimental limits from LEP2 on MSSM Higgs bosons are significant.
For maximal mixing ($X_t\equiv A_t-\mu\cot\beta=\sqrt 6 \mstop$),
one finds that $\mhl,\mha\lsim 91\gev$ \cite{lephiggs}, implying that
$\tanb\lsim 2.5$ is excluded.
But, this analysis assumes $\mstop<1\tev$, a CPC
Higgs sector, and the absence of invisible decays.
For higher $\mstop$, the value of $\mhl$
predicted for a given $[\mha,\tanb]$ choice
increases and less of the parameter space is excluded.
Allowing for CP-violation weakens the lower bounds on the MSSM
Higgs boson mixed states and the lower bound 
on $\tanb$~\cite{Kane:2000aq,Carena:2000ks}.
Allowing for the $\hl$ and $\ha$ to have  
substantial invisible decays might substantially 
weaken the constraints on the $\hl\ha$ cross section. The $\epem\to ZX$
channel would have to be relied upon much more heavily.
A LEP2 study of this scenario would be worthwhile.

\begin{figure}
\begin{center}\includegraphics[width=6in]{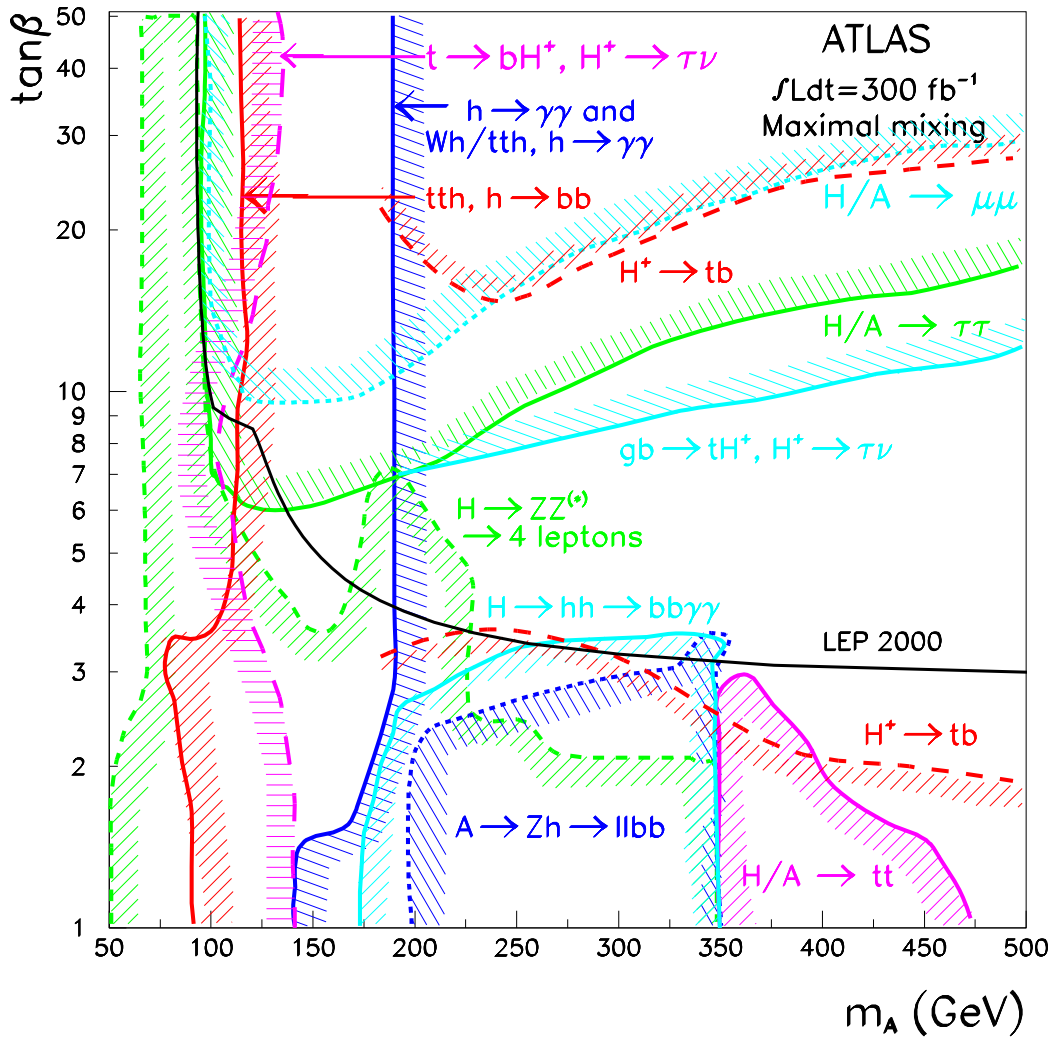}
\caption{\label{atlasfig}
$5\sigma$ discovery contours for MSSM Higgs boson detection at the LHC
in various channels are shown in the $[\mha,\tanb]$ parameter plane, 
assuming maximal mixing and an integrated luminosity of $L=300\fbi$
for the ATLAS detector. This figure is preliminary.
}
\end{center}
\end{figure}

Prospects for discovery of at least one MSSM Higgs boson at
future colliders are excellent.  Our discussion will focus
on the `decoupling' limit of $\mha>2\mz$ in which the $\hl$
is quite SM-like with full $VV$ coupling strength. At the Tevatron,
$5\sigma$ discovery of the $\hl$
will be possible in $q'\bar q \to V \hl$ ($V=W,Z$, $\hl\to b\anti b$)
with $L>20-35\fbi$ of accumulated luminosity, the larger $L$'s being
required for higher $\mha$, depending upon the SUSY-breaking
scenario. For $\tanb= 20-50$ and $\mha=100-200\gev$ (the higher $\tanb$
values being required for the higher $\mha$ values),
the $\hh,\ha$ can be discovered at the Tevatron in
$gg,q\bar q\to b\bar b\hh,b\bar b \ha\to 
b\anti b b\anti b$~\cite{Carena:1999gk,Carena:2000bh}.  
(The $\hh,\ha\to \tauptaum$ decay modes have not
been studied yet.) Given that $\mha>2\mz$ is a very common implication of
RGE induced EWSB scenarios, it is most likely that only the $\hl$
will be seen at the Tevatron. 
The LHC has somewhat greater sensitivity to the full complement of MSSM
Higgs bosons. The best signals for the $\hl$ are expected
in the $gg\to\hl\to \gam\gam$, $t\anti t\hl\to t\anti t b\anti b$
and $WW\to \hl\to\tauptaum$ channels. (Large rates for $\hl$
production in the decay chains of heavy SUSY particles, such
as the gluinos and squarks, are also possible in some scenarios
and would produce dramatic signals.)
At high enough $\tanb$ values, the
$gg,q\bar q\to b\bar b\hh,b\bar b \ha$, with $\hh,\ha\to
\tauptaum$ or $\mupmum$ and $gb\to \hpm t$ with 
$\hpm\to \tau^\pm \nu$ will provide good signals for the heavier Higgs bosons.
These signals have been studied by both the ATLAS and CMS collaborations.
The exact reach of these primary channels is illustrated by the ATLAS
results \cite{atlasmaxmix} of Fig.~\ref{atlasfig}.
Note that for maximal mixing,
LEP2 limits exclude $\tanb\lsim 2.5$, the region
where other MSSM Higgs discovery modes could be important at the LHC.
For $\tanb\gsim 3$ but below 8 to 15 
(for $\mha=250\gev$ to $500\gev$, respectively), 
there is a wedge of parameter space in
which only the $\hl$ will be detected. An important question then
becomes whether a high energy linear collider or a possible future muon
collider could detect the $\hh,\ha$ in this wedge region of parameter
space, or at least give some indication of the value of $\mha\sim \mhh\sim
\mhpm$ if their masses are large.

Discovery of the $\hl$ will be straightforward at a LC, using
the same production/decay modes as for a light $\hsm$.
The high rates imply that
precision measurements of the couplings of the $\hl$ will be possible,
possibly allowing the detection of deviations from expectations
for the $\hsm$ even when $\mha$ is fairly 
large \cite{Gunion:1997qz,Carena:2001bg}.
In the simpler SUSY-breaking scenarios (\eg\ maximal mixing or minimal mixing),
detection of such deviations will be possible for $\mha\lsim 500-600\gev$
for $L=1\abi$ of integrated luminosity at $\rts\sim 500\gev$,
and would provide a crucial indication of where in mass to search for
the $\hh$, $\ha$ and $\hpm$.
This will be particularly important if $\mha+\mhh,2\mhpm>2\rts$
(so that $\hh\ha$ and $\hp\hm$ pair production is impossible at the LC)
and if, in addition, $[\mha,\tanb]$ lie in the LHC no-discovery wedge.
Analogous to our discussion for a decoupled $\h$ of a general 2HDM,
very high $\tanb$ is required for an observable signal in
the $\epem\to b\anti b \ha,b\anti b \hh, bt\hpm$ channels.
If SUSY particles and a light SM-like $\hl$ are detected, 
even if the $\hh,\ha,\hpm$ are not detected 
we will be quite certain that a set of heavier Higgs bosons must exist.
The challenge is to zero-in on colliders/techniques for discovering them.

\begin{figure}
\begin{center}
\includegraphics[width=6in]{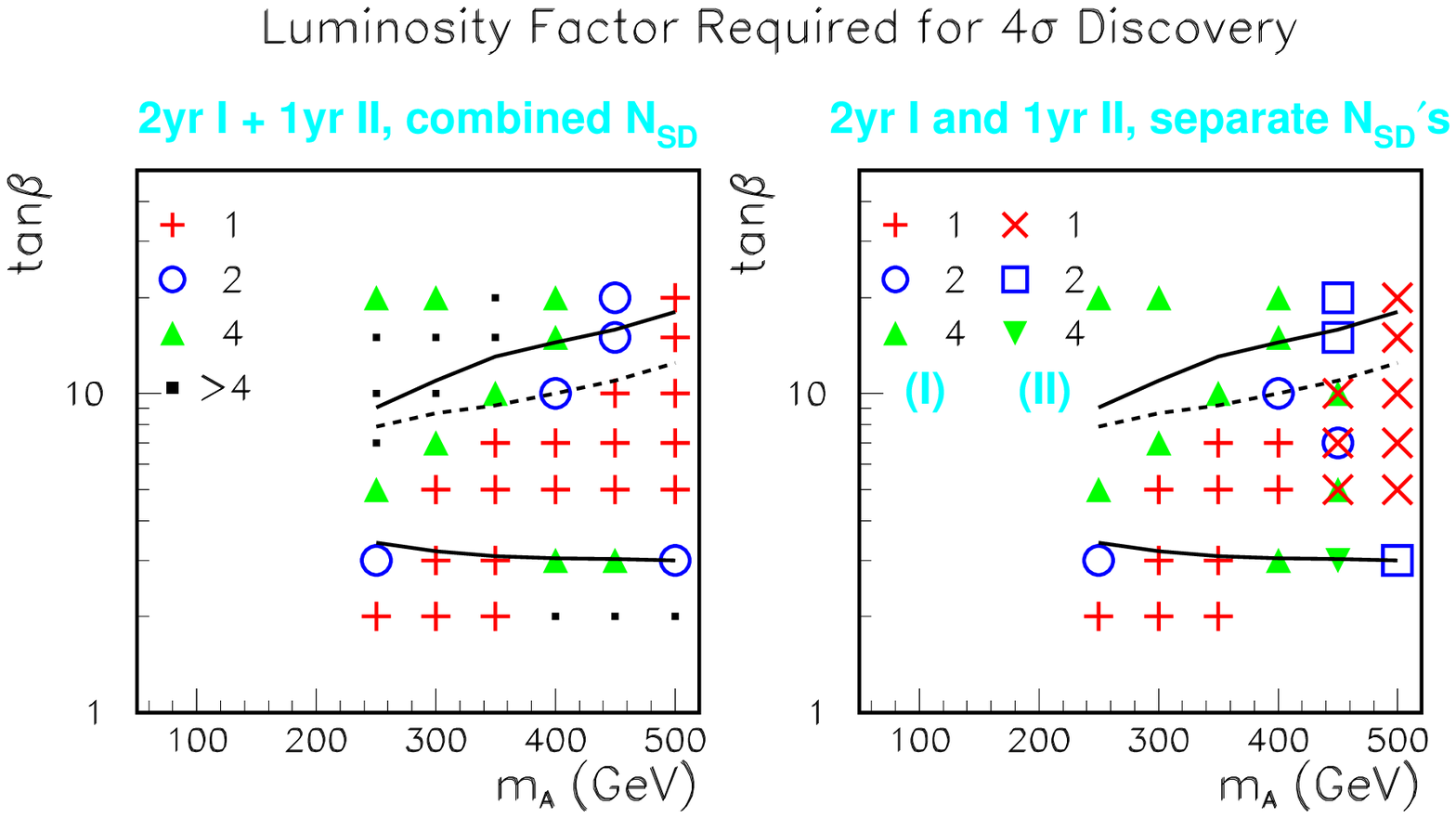} 
\vspace*{-3in}
\caption{
Assuming a machine energy of $\rts=630\gev$,
we show the $[\mha,\tanb]$ points for which two $10^7$ sec years
of running using a broad $E_{\gam\gam}$ spectrum (I)  
and one $10^7$ sec year of running using 
a spectrum peaked at $E_{\gam\gam}\sim 500\gev$ (II)
will yield $S/\sqrt{B}\geq4$.  In the left-hand window we
have combined results from the type-I and type-II running using
$S/\sqrt B=\sqrt{ S_I^2/B_I+S_{II}^2/B_{II}}$. In the right-hand
window, we show the separate results for $S_I/\sqrt{B_I}$ and 
$S_{II}/\sqrt{B_{II}}$. 
The solid curves indicate the wedge region from the LHC plot
of Fig.~\ref{atlasfig} --- the lower black curve is that from the
LEP (maximal-mixing) limits, but is
somewhat higher than that currently claimed by the LEPEWWG,
while the upper solid curve is that above which $\hh,\ha\to\tauptaum$
can be directly detected at the LHC.  For parameter choices above
the dashed curve, $\hpm\to\tau^\pm\nu_\tau$ can be directly detected
at the LHC. Also shown are the additional points for
which a $4\sigma$ signal level is achieved if the total
luminosity is doubled or quadrupled (the `2' and `4' symbol cases)
relative to the one-year luminosities we are employing. 
(The small black squares in the LH window indicate the additional
points sampled for which even a luminosity increase of a factor
of 4 for both types of running does not yield a $4\sigma$ signal.)
Such luminosity
increases could be achieved for some combination of longer running time and/or
improved technical designs. For example, the factor of `2' 
results probably roughly apply to TESLA. \label{gamgammssm}   
}
\end{center}
\end{figure}

In this regard, production of these heavy Higgs bosons
in the $s$-channel at both $\gam\gam$ \cite{Gunion:1993ce}
and $\mupmum$ \cite{Barger:1997jm} colliders 
could provide detectable signals.
If we have some indication of the value of $\mha$ (\eg\ from
detection of $\hl$ vs. $\hsm$ deviations), then we will
know exactly what energy to employ. The expectations
for a $\gam\gam$ collider are explored in some detail 
in~\cite{Muhlleitner:2001kw,gunasner}.
If $\mha$ is known within $\sim 50\gev$, less than one year
of operation of the $\gam\gam$ collider with $E_{\gam\gam}$
luminosity peaked at $E_{\gam\gam}\sim \mha$ will be needed to detect
the $\hh,\ha$ signal.
But, if the indirect determination of $\mha$ is believed to
be unreliable, or the SUSY scenario is such that no deviations
will be present regardless of the value of $\mha$, one
must employ a different strategy. One possibility
is $\gam\gam$ collisions for LC operation at maximum energy,
presumed in \cite{gunasner} to be $\rts=630\gev$ so as to
allow substantial luminosity for $E_{\gam\gam}$ up to $500\gev$.
By running for two years with laser and electron
polarizations and orientation such as to yield a broad $E_{\gam\gam}$ spectrum
and for one year with the $E_{\gam\gam}$ spectrum peaked at $500\gev$,
detection of the $\gam\gam\to\hh,\ha\to b\anti b$ signal will
be possible throughout much of the LHC no-discovery wedge region. 
This is illustrated in Fig.~\ref{gamgammssm}.
The results shown assume that 50\% of the $\hh,\ha$ signal events
will fall into a single $m_{b\anti b}$ mass bin of size 10 GeV,
as consistent with expected mass resolutions and predicted Higgs widths.

Because of lack of space, I only summarize expectations for
a muon collider Higgs factory with energy in the $250-500\gev$
range. A $4\sigma$ $\mupmum\to\hh,\ha\to b\anti b$ signal could be
found, either using operation at $\rts\sim 500\gev$ and the
bremsstrahlung  (radiative return) tail or by employing
an appropriate scan strategy, for almost all values of $[\mha,\tanb]$
in the LHC wedge region \cite{Barger:1997jm,jfgucla,bbghsm01}.

Of course, there are variants of these `standard' results 
that temper this relatively optimistic outlook.
\bit
\item
Invisible decays will probably 
allow non-detection scenarios at hadron colliders.
This is important even for the $\hl$. Indeed,
$\hl\to\cnone\cnone$ is still possible given the LEP2 data. 
For large $\br(\hl\to\cnone\cnone)$, it is necessary to
have  $\mcnone/\mcpmone\sim M_1/M_2<1/2$, 
\ie\ smaller than predicted by universal
boundary conditions, in order that $\hl\to \cnone\cnone$ not be kinematically
suppressed given the $\mcpmone>103\gev$ lower limit from LEP2.
In the study of~\cite{Belanger:2001am,Belanger:2000tg} 
(see also \cite{Griest:1988qv,Djouadi:1996mj}),
it is found that the universal boundary condition
prediction of $M_1/M_2=1/2$ allows for at most
$\br(\hl\to\cnone\cnone)\sim 20\%$, whereas
$M_1/M_2=1/10-1/5$ allows $\br(\hl\to\cnone\cnone)>50\%$.
One also needs substantial $\hl\to\cnone\cnone$ coupling, which in
turn requires that the
$\cnone$ have substantial higgsino content.  This latter is possible
when $\mu$ (and $M_2$) are not large. 
(One should recall that small $M_1$, $M_2$ and $\mu$ are preferred by current
results for $a_\mu$.)
\item
Similarly, the usual LHC contours for $\hh,\ha,\hpm$ discovery in
various modes will be modified (at low to moderate $\tanb$
when $\mha>2\mz$) if $\cnone\cnone,\cpone\cmone,\stau^+\stau^-,\snu\anti{\snu},
\ldots$ decays are kinematically allowed \cite{Gunion:1988yc,hhg,Djouadi:1999xd}.
However, at high $\tanb$ the usual dominance of decays to $b\anti b$ and
$\tauptaum$ will be preserved. This, implies that even if SUSY particles
are light the widening of the $\hl$-only LHC wedge at high $\tanb$
will be moderate (and the LEP2 limits mean that we do not need
to worry very much about low $\tanb$).
\item
Stop loop correction to $gg$ and $\gam\gam$ couplings of the
MSSM Higgs bosons can be substantial~\cite{hhg,Djouadi:1998az}. In particular, 
stop and top loop contributions to $gg$ fusion negatively interfere, implying
some reduction of $gg$ fusion production of the $\hl$ when stops
are light, but also
some increase in $\br(\hl\to\gam\gam)$.
\item 
Radiative corrections to Higgs couplings can result
in early or even exact decoupling,
\ie\ $\cos^2(\beta-\alpha)=0$ independent of $\mha$.
\item
Radiative corrections 
can also greatly modify expectations for  
$\hl\to b\anti b$ decays~\cite{Carena:2001bg}.
The important loops here do not decouple when SUSY masses are large.
In one extreme, for special, but not unreasonable, parameter choices, one finds
$\hl\sim H_u$, where $H_u$ is the MSSM doublet field that
couples to up quarks (only), and $\br(\hl\to b\anti b)\sim 0$.
In another extreme, substantial enhancement of the $\hl\to b\anti b$
coupling occurs. 

In either case, there are many implications for
$\hl$ discovery. For example, suppressed 
$\Gamma(\hl\to b\anti b)$ implies enhanced $\br(\hl\to \gam\gam),
\br(\hl\to WW^*)$, and it is even possible that
detection of the $\hl$ in its $\gam\gam$ decay mode could be possible
at the Tevatron for some range of
$\mhl$ if $\hl\sim H_u$ \cite{Mrenna:2001qh}.
Since $\lam_b$ can be either enhanced or suppressed, it is useful to note
\cite{Carena:2000bh} that the
LHC $gg\to \hl\to\gam\gam$ and Tevatron $W\hl[\to WW^*]$ 
modes improve when the LHC, Tevatron $W,Z\hl[\to b\anti b]$ modes deteriorate.
There is also complementarity between the Tevatron and LHC in that
as the $b\anti b\hl$ coupling and $\mhl$ vary one finds  
that $\hl$ discovery will occur at one or the other machine,
even if not at both. 

\eit
Of course, at the LC the $\epem\to Z\hl$ mode is robust regardless
of how the $\hl$ decays. Further, at the LC,
$\hh\ha$ and $\hp\hm$ detection are quite robust against complicated
decays if pair production is not too near the kinematic limit \cite{Gunion:1997cc,Gunion:1996qd,Feng:1997xv}.
In fact, the precise decay mixtures provide an immensely powerful probe
of the soft SUSY breaking parameters. It is only necessary to separate
different final state channels 
([$3\ell,2b$], [$1\ell,0b]$, . . . . --- maybe 15 or 20 different
channels) from one another and have precise knowledge of the 
efficiencies for different channels.

The above discussion was restricted to the MSSM. There is good
reason to suppose that the Higgs sector could have one or more
singlets beyond the required two-doublets.  Singlet Higgs fields do
not disturb coupling constant unification and lead to some very
attractive improvements to the MSSM.
The simplest model is the NMSSM in which
a single Higgs singlet is introduced \cite{Ellis:1989er}. (See \cite{hhg}
for a review and further details.)
The new attractive feature of this model
is that the superpotential can contain
the term $W\ni \lam \hat H_1 \hat H_2 \hat N$,
such that for $\vev N\neq 0$ there is a natural source
and appropriate magnitude, $\lam \vev N=\mu$,
for the somewhat mysterious $\mu \hat H_1\hat H_2$
superpotential term of the MSSM.
In the NMSSM, there are three CP-even Higgs bosons ($h_{1,2,3}$) and 
two CP-odd Higgs bosons ($a_{1,2}$), assuming no CP violation.
As we have already discussed, we can add any number of singlets and
still find a Higgs boson signal  for 
 $\epem\to Z^*\to Z h_i$ production at a LC, even if the signals overlap.
At the LHC, establishing a corresponding guarantee is quite challenging.
Indeed, it was shown in \cite{Gunion:1996fb}
that parameters of the NMSSM could be chosen
so that no Higgs boson would be detected in the 
modes for which definitive experimental
results were available at the time of the Snowmass 1996 workshop.
The modes employed in 1996 were:
1) $\zstar\to Z\h$ at LEP2; 
2) $\zstar\to \h\a$ at LEP2;
3) $gg\to \h\to\gam\gam$ at the LHC; 
4) $gg\to\h\to Z\zstar~{\rm or}~ZZ\to 4\ell$
at the LHC; 
5) $t\to\hp b$ at the LHC;
6) $gg\to b\anti b \h,b\anti b\a \to b\anti b \tauptaum$ at LHC;
7)  $gg\to\h,\a\to\tauptaum$ at the LHC.
The regions of parameter space in which no Higgs bosons
would be detected were characterized by substantial
mixing among all the Higgs bosons and moderate $\tanb$ values.
This study has been updated as part of the Snowmass01
and LesHouches01 workshops~\cite{nmssmnew}.
One important discovery mode not confirmed by the experimental groups
at the time of Snowmass96 is 
$tt\to t\anti t h_i\to t\anti t b\anti b$ \cite{Dai:1993gm}.
The experimental groups now believe that this 
will be visible~\cite{Richter-Was:1999sa,Sapinski:2001tn} if the $h_i$
coupling to $t\anti t$ is comparable to the $\hsm t\anti t$ coupling.
In \cite{nmssmnew}, the full NMSSM
parameter space (excluding regions for which SUSY pairs
or Higgs bosons appear in Higgs decays) was rescanned
including the $t\anti t \h$ mode with the result that most 
(but not all) parameter choices
for which Higgs discovery would not have been possible in the 1996 analysis
{\it would} lead to one of the Higgs bosons being
visible in this mode.  In addition, we find
(using the theoretical estimates of \cite{Zeppenfeld:2000td})
that essentially all of 
the remaining `bad' portions of parameter space would lead
to visible signals in the modes where one of the $h_i$
is produced via $WW$-fusion and then decays to $\tauptaum$.
This illustrates the great importance that the ATLAS and CMS
groups should attach to further improving their Higgs
discovery techniques, particularly by adding new 
modes complementary to those already considered.

\section{Determining the CP of an observed Higgs boson}

Determination of the CP properties
of the Higgs bosons could prove very crucial to
sorting out a complex Higgs sector.
At a LC there are many techniques based on $WW$
and/or $ZZ$ couplings for verifying a substantial
CP=+ component. But the $VV$ couplings are
only sensitive to the CP=$-$ component of a Higgs boson at one-loop level.
As a result, using such couplings it is 
very hard to see a CP=$-$ coupling even if it is present.
Since CP=+ and CP=$-$ couplings to $t\anti t$ of any $\h$
are both tree-level ($\anti t (a+ib\gamma_5) t$,
where $a$,$b$  is the CP-even, -odd Higgs component), 
angular distributions of the $t$, $\anti t$
and $\h$ relative to one another in the $t\anti t\h$ final 
state allow determination of the
relative sizes of $a$ and $b$ for lighter $\h$'s \cite{Gunion:1996xu}.  
The best approach is to use the optimal observable 
technique~\cite{Gunion:1996vv}.
At a LC, as long as there is reasonable event rate
(which requires $\rts>800\gev$ for $\mh\sim 100-200\gev$)
this is straightforward~\cite{Gunion:1996vv}.
At the LHC, there will be a high event rate, 
but reconstruction and identification of the $t$ and $\anti t$ is trickier and
backgrounds will be larger. Still, 
there is considerable promise~\cite{Gunion:1996xu,Gunion:1998hm}.

The CP=+ and CP=$-$ components of a Higgs boson
also couple with similar {\it magnitude}
 but different structure to $\gam\gam$
(via 1-loop diagrams). Thus, determination of the CP properties of any
Higgs boson that can be seen in $\gam\gam$ collisions at the LC
will be possible~\cite{Grzadkowski:1992sa,Kramer:1994jn,Gunion:1994wy}
by comparing production rates for different orientations of the
polarizations of the colliding $\gam$'s.
Very briefly, we have
\def\cala{{\cal A}}
\beq
\cala_{CP=+}\propto \vec \eps_1\cdot\vec \eps_2\,,\quad
\cala_{CP=-}\propto (\vec\eps_1\times\vec \eps_2)\cdot \hat p_{\rm beam}\,.
\eeq
For pure CP states, one will want to
maximize the linear polarization of the back-scattered photons
and adjust the orientation
($\perp$ for CP odd dominance, $\parallel$ for CP even
dominance) to determine the CP nature of the Higgs boson being produced.
For a light SM-like Higgs boson, a detailed study~\cite{gunasner}
can determine that ${\cal CP}=+1$ with an error of $\delta{\cal CP}/{\cal CP}\sim 0.11$. One can also separate the
$\ha$ from the $\hh$ when these are closely degenerate (as typical for
$\tanb\gsim 4$ and $\mha>2\mz$).
For mixed CP states, one achieves better statistics by
using circularly polarized
photons and employing
helicity asymmetries to determine the CP mixture.

At a muon collider Higgs factory there is a particularly appealing
approach using asymmetries involving transversely 
polarized muon beams \cite{Barger:1997jm,Atwood:1995uc}. 
For resonance, $R$, production with 
$\anti\mu(a+ib\gamma_5)\mu$ coupling to the muon, one has 
\bea
\overline\sigma_S(\zeta)&=&\overline\sigma_S^0\left(
1+P_L^+P_L^-+P_T^+P_T^-\left[{a^2-b^2\over a^2+b^2}\cos\zeta-{2ab\over a^2+b^2}\sin\zeta \right]
\right)\,,
\label{sigform}
\eea
where $P_T$ ($P_L$) is the degree of transverse (longitudinal)
polarization of the colliding $\mu^+$ and $\mu^-$, and $\zeta$ is the 
angle of the $\mu^+$ transverse polarization relative
to that of the $\mu^-$ as measured using the the direction of the $\mu^-$'s
momentum as the $\hat z$ axis.
Only the $\sin\zeta$ term is truly CP-violating, but the
dependence on $\cos\zeta$ is also
sensitive to $a/b$.  One must take into account the precession
of the $\mu^+$ and $\mu^-$ as they circulate around the storage ring.
Fortunately, this is easy to do and very decent accuracy is possible
for the determination of $b/a$ for a Higgs boson after a few years of 
operation~\cite{Grzadkowski:2000hm},
provided the Higgs factory can achieve luminosities
about a factor of two larger than the current benchmarks.

\section{Conclusions}

There are a large variety of very viable Higgs sector models.
Experiment will be required to determine the correct theory.
The current and future machines
and the related tools and techniques 
that have been developed have reached a high enough
level of sophistication that we should have
a good chance of detecting and studying
the Higgs bosons of even rather unusual Higgs sectors.

% figures should be put into the text as floats.
% Use the graphicx package (distributed with LaTeX2e).
% See the LaTeX Graphics Companion by Michel Goosens, Sebastian Rahtz,
% \begin{figure}
% \caption{}
% \label{}
% \end{figure}

% tables follow here or maybe be put in the text
% \begin{table}
% \caption{}
% \label{}
% \end{tabular}
% \end{table}

%\begin{acknowledgments}
%\end{acknowledgments}

%\bibliography{your bib file}

\end{document}